\def\lJ{{\lambda_{_{J}}}}
\def\nL{{\mathcal{L}}}
\def\nW{{\mathcal{W}}}

\documentclass[aps,prb,twocolumn,groupedaddress,showkeys,showpacs,superscriptaddress,floatfix]{revtex4-1}
\usepackage{epsfig}
\usepackage{multirow}
\usepackage{amsmath, amssymb,mathtools}
\usepackage{color}
\usepackage{graphicx}% Include figure files
\usepackage{dcolumn}% Align table columns on decimal point
\usepackage{bm}% bold math
\usepackage{subfigure}
\usepackage[utf8]{inputenc}
\usepackage[T1]{fontenc}

\DeclarePairedDelimiter{\abs}{\lvert}{\rvert}

\begin{document}

\title{\bf Coherent diffraction of thermal currents in long Josephson tunnel junctions}

\author{Claudio Guarcello\thanks{e-mail: claudio.guarcello@nano.cnr.it}}
\affiliation{SPIN-CNR, Via Dodecaneso 33, 16146 Genova, Italy}
\affiliation{NEST, Istituto Nanoscienze-CNR and Scuola Normale Superiore, Piazza S. Silvestro 12, I-56127 Pisa, Italy}
\affiliation{Radiophysics Department, Lobachevsky State University, Gagarin Ave. 23, 603950 Nizhni Novgorod, Russia}
\author{Francesco Giazotto\thanks{e-mail: giazotto@sns.it}}
\affiliation{NEST, Istituto Nanoscienze-CNR and Scuola Normale Superiore, Piazza S. Silvestro 12, I-56127 Pisa, Italy}
\author{Paolo Solinas\thanks{e-mail: paolo.solinas@spin.cnr.it }}
\affiliation{SPIN-CNR, Via Dodecaneso 33, 16146 Genova, Italy}

\date{\today}

\begin{abstract}
We discuss heat transport in thermally-biased long Josephson tunnel junctions in the presence of an in-plane magnetic field. In full analogy with the Josephson critical current, the phase-dependent component of the heat current through the junction displays coherent diffraction. Thermal transport is analyzed as a function of both the length and the damping of the junction, highlighting deviations from the standard “Fraunhofer” pattern characteristic of short junctions. The heat current diffraction patterns show features strongly related to the formation and penetration of Josephson vortices, i.e., solitons. 
We show that a dynamical treatment of the system is crucial for the realistic description of the Josephson junction, and it leads to peculiar results.
In fact, hysteretic behaviors in the diffraction patterns when the field is swept up and down are observed, corresponding to the trapping of vortices in the junction.
\end{abstract}

\pacs{85.25.Cp, 74.50.+r, 74.25.Sv, 74.78.Na}

\maketitle

\section{Introduction}
\label{Intro}\vskip-0.2cm

The signature of phase coherent heat currents in an extended Josephson junction (JJ) was recently successfully confirmed by means of diffraction patterns experiments~\cite{Mar14}. Specifically, the Fraunhofer diffraction for thermal currents manifests itself with a modulation of the electron temperature in a small metallic electrode nearby contacted to a temperature-biased short JJ when sweeping the external magnetic field. 
The proof of the phase-coherent behavior in thermally-biased JJs paved the way to the implementation of superconducting hybrid coherent caloritronic~\cite{MarSol14} circuits such as, for instance, interferometers~\cite{Gia12,Mar14,For16}, heat diodes~\cite{For14,Mar15} and transistors~\cite{DAm15,For16}, and solid-state memory devices~\cite{Xie11}.

In this paper we theoretically investigate heat transport in temperature-biased extended \emph{long} JJs, showing interference of the phase-dependent component of thermal current in the presence of an in-plane magnetic field. Accordingly, a heat diffraction pattern results, in full analogy to what occurs for the Josephson critical current. We discuss the influence of the length and the damping of the device on both the characteristics of the diffractions patterns and the configurations of Josephson vortices, i.e., solitons, set along the junction. Moreover, according to a full dynamical description of the system, hysteretic behaviors of the critical heat current and the magnetic flux through the JJ come to light. 

To measure the thermal effect we discuss in long JJ a set-up similar to the one recently proposed~\cite{Gia13} and successfully implemented~\cite{Mar14} for a thermally biased rectangular \emph{short} JJ can be used. Specifically, this junction consists of a first electrode coupled to two source and drain normal metal electrodes, allowing Joule eating and thermometry, and a second one, extending into a large bonding pad that is kept open during the heat diffraction experiment. An extra probe is connected, through a bias JJ, to the first electrodes to perform the electric characterization of the device. 

The paper is organized as follows. In Sec.~\ref{Model}, the model used to describe a thermal biased and electrically open long JJ is shown. Specifically, first we look at the phase dependence of the maximum heat current and then we introduce the mathematical approach used to describe the phase dynamics of an electrically open long JJ in the presence of an external uniform magnetic field. In Sec.~\ref{Results} the results are shown and analyzed, by focusing on the thermodynamical picture of the problem and its limits when the effects of the damping are taken into account, the emerging hysteretic behavior as a function of the magnetic field, and the effect of the thermal fluctuations. In Sec.~\ref{Conclusions} conclusions are drawn.

\section{The Model}
\label{Model}\vskip-0.2cm 
\begin{figure}[b!!]
\centering
\includegraphics[width=0.5\textwidth]{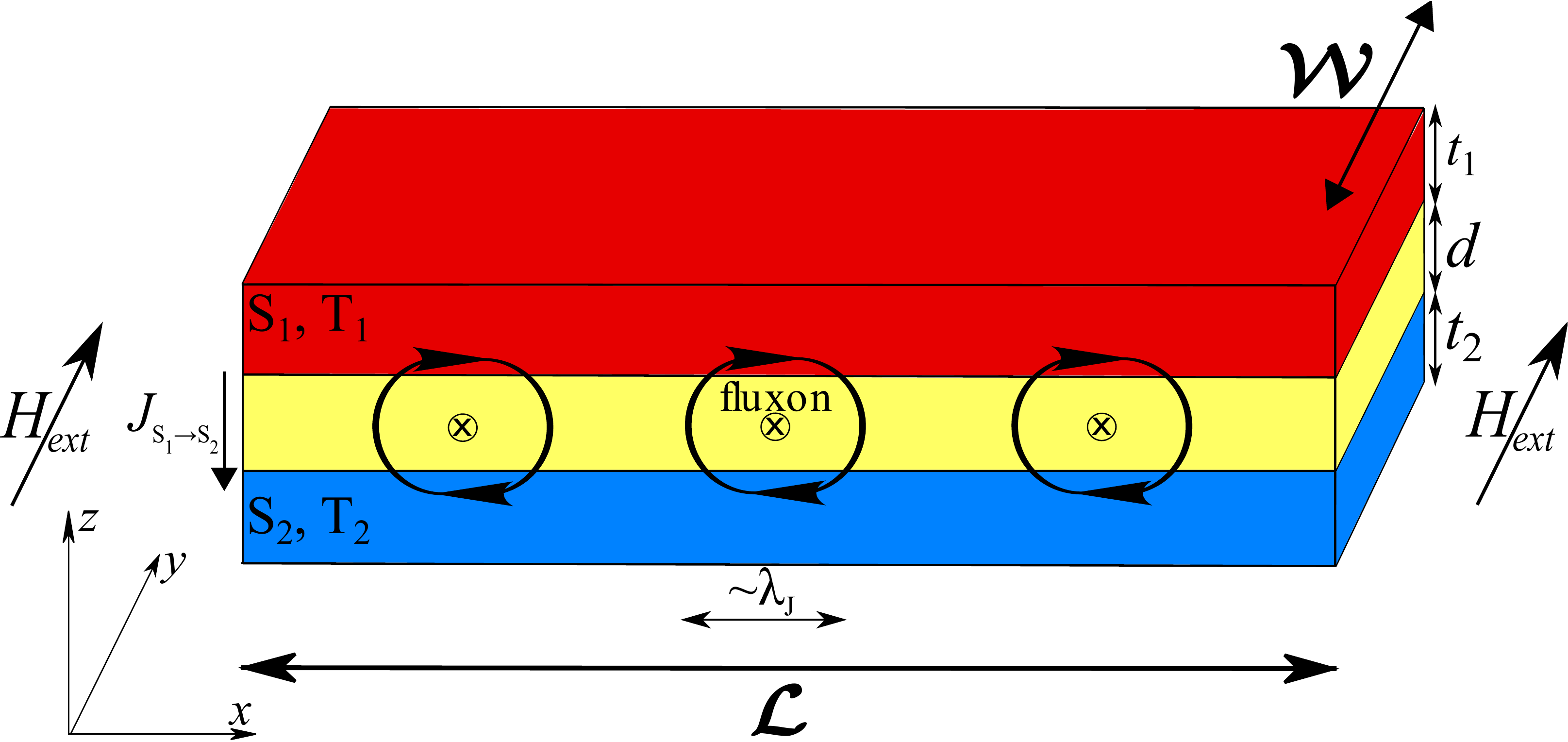}
\caption{(Color online) Temperature biased long Josephson junction. The heat current $J_{S_1\rightarrow S_2}$ flows along the $z$ direction whereas a constant, homogeneous external magnetic field $H_{ext}$ is applied in the $y$ direction. The length and the width of the junction are $\nL>\lJ$ and $\nW\ll \lJ$, respectively, where $\lJ$ is the Josephson penetration depth. $T_i$ and $t_i$ represent the temperature and the thickness of the superconductor $S_i$, respectively, and $d$ is the interlayer thickness. Fluxons in the junction, extending over a physical distance of the order of $\lJ$, are also represented. }
\label{Fig1}
\end{figure}

The system analyzed is a symmetric extended long JJ with dimensions $\nL$ and $\nW$, shown in Fig.~\ref{Fig1}, formed by two superconducting electrodes $S_1$ and $S_2$ in a thermal steady state residing at different temperatures $T_1$ and $T_2$, respectively, with $T_1\ge T_2$. In the long junction limit, the lateral dimension $\nL$ is greater than the Josephson penetration depth~\cite{Bar82} 
\begin{equation}
\lJ=\sqrt{\frac{ \Phi_0}{2\pi\mu_0i_ct_d}}, 
\label{LambdaJ}
\end{equation}
specifically $\nL>\lJ$ and $\nW\ll\lJ$, where $\Phi_0=h/2e\simeq2.067\times10^{-15} \textup{Wb}$ is the magnetic flux quantum, $\mu_0$ is the vacuum permeability, $i_c$ is the critical current area density, and $t_d=\lambda_1(T_1)+\lambda_2(T_2)+d$ is the effective magnetic thickness, $\lambda_i(T_i)=\lambda_i(0)\Big/\sqrt{1-\left ( T_i/T^i_c \right )^4}$ being the London penetration depth of the superconductor $S_i$ and $d$ is the interlayer thickness ($T^i_c$ is the critical temperature of the superconductor $S_i$). If $\lambda_i>t_i$ the effective magnetic thickness has to be replaced by $\tilde{t}_d=\lambda_1\tanh\left ( t_1/2\lambda_1 \right )+\lambda_2\tanh\left ( t_2/2\lambda_2 \right )+d$. The magnetic field $H_{ext}$ lies parallel to a symmetry axes of the junction and along $y$.

In the presence of a temperature gradient and with no voltage bias, a finite heat current $J_{S_1\rightarrow S_2}$, see Fig~\ref{Fig1}, flows through the junction from $S_1$ to $S_2$~\cite{Mak65,Gut97,Gut98,Zha03,Zha04,Gol13}
\begin{equation}
 J_{S_1\rightarrow S_2}\left ( T_1,T_2,\varphi \right )\text{=}J_{qp}\left ( T_1,T_2 \right )-J_{int}\left ( T_1,T_2 \right )\cos\varphi
\label{TotalThermalCurrent}
\end{equation}
where $\varphi$ is the macroscopic quantum phase difference between the superconductors.
In Eq.~(\ref{TotalThermalCurrent}), $J_{qp}$ is the heat flux carried by quasiparticles~\cite{Gia06,Fra97} and $J_{int}$ is the phase-dependent part of the heat current which is peculiar to JJs. The latter originates from the energy-carrying processes involving Cooper pairs tunneling and recombination or destruction of Cooper pairs. Since the phase difference between the annihilated and created pairs is relevant in such a process this gives rise to the $\cos\varphi$ contribution to the transferred heat. 
The oscillatory behavior of the thermal current $J_{S_1\rightarrow S_2}$ was experimentally verified in Ref.~\onlinecite{Gia12,Mar14}.

The first term on the rhs of Eq.~(\ref{TotalThermalCurrent}) explicitly reads~\cite{Mak65,Gut97,Gut98,Zha03,Zha04,Gol13}
\begin{small}
\begin{equation}
J_{qp}\text{=}\frac{1}{e^2R_N}\underset{0}{\overset{\infty}{\mathop \int }}d\varepsilon \varepsilon \mathcal{N}_1 ( \varepsilon ,T_1 )\mathcal{N}_2 ( \varepsilon ,T_2 ) [ f ( \varepsilon ,T_2 ) -f ( \varepsilon ,T_1 ) ]
\end{equation}
\end{small}%
where
\begin{equation}
\mathcal{N}_i\left ( \varepsilon ,T_i \right )=\frac{\left | \varepsilon \right |}{\sqrt{\varepsilon ^2-\Delta _i\left ( T_i \right )^2}}\Theta \left [ \varepsilon ^2-\Delta _i\left ( T_i \right )^2 \right ]
\end{equation}
is the BCS normalized density of states in $S_i$ at temperature $T_i$ ($i=1,2$). Here $\varepsilon$ is the energy measured from the condensate chemical potential, $\Delta _i\left ( T_i \right )$ is the temperature-dependent superconducting energy gap, $f\left ( \varepsilon ,T_i \right )=\tanh\left ( \epsilon/2 k_B T_i \right )$, $\Theta(x)$ is the Heaviside step function, $k_B$ is the Boltzmann constant, $R_N$ is the junction normal-state resistance, and $e$ is the electron charge. The second term on the rhs of Eq.~(\ref{TotalThermalCurrent}) reads~\cite{Mak65,Gut97,Gut98,Zha03,Zha04,Gol13}
\begin{small}
\begin{equation}
J_{int}\text{=}\frac{1}{e^2R_N}\underset{0}{\overset{\infty}{\mathop \int }}d\varepsilon \varepsilon \mathcal{M}_1 ( \varepsilon ,T_1 )\mathcal{M}_2 ( \varepsilon ,T_2) [ f ( \varepsilon ,T_2 ) - f ( \varepsilon ,T_1 ) ]
\end{equation}
\end{small}%
where
\begin{equation}
\mathcal{M}_i\left ( \varepsilon ,T_i \right )=\frac{\Delta _i\left ( T_i \right )}{\sqrt{\varepsilon ^2-\Delta _i\left ( T_i \right )^2}}\Theta \left [ \varepsilon ^2-\Delta _i\left ( T_i \right )^2 \right ].
\end{equation}
is the Cooper pair BCS density of states in $S_i$ at temperature $T_i$~\cite{Bar82}.
The $\varphi$-dependent component of the heat current can be expressed as
\begin{equation}
J_H ( T_1, T_2, H )\text{=}\iint dx dy J_A(x,y,T_1, T_2)\cos \varphi(x,y) 
\label{PhiThermalCurrent}
\end{equation}
where $J_A(x,y,T_1, T_2)$ is the heat current density per unit area. By supposing a spatially uniform heat current density, $J_{int}\left ( T_1, T_2 \right ) = \nW \nL J_A\left ( T_1, T_2 \right )$.

In Eq.~(\ref{PhiThermalCurrent}), $\varphi(x,y)$ is the phase difference induced by the applied magnetic field $H_{ext}$. In fact, $\varphi(x,y)$ depends on the normalized local magnetic field $h_y(x)$ through the equations~\cite{Bar82}
\begin{equation}
\frac{\partial \varphi }{\partial x}=\frac{2\pi\mu_0t_d}{\Phi_0}H_y(x)=h_y(x) \qquad \qquad \frac{\partial \varphi }{\partial y}=0.
\label{LocalMagneticField}
\end{equation}
The latter equation comes from the condition $\nW\ll \lJ$, so that $\varphi \left ( x,y \right )\equiv \varphi \left ( x \right )$.
For a short rectangular JJ the external magnetic field is spatially homogeneous along the junction, namely $H_y(x)\equiv H_{ext}$, so that the phase is linearly increasing, $\varphi(x)=2\pi\mu_0t_d/\Phi_0 H_{ext} x+\varphi_0$. Instead, in a long JJ both the penetrating external field and the self-field generated by the Josephson current have to be considered, so that $\varphi(x)$ nonlinearly changes along the junction.
Therefore, Eq.~(\ref{PhiThermalCurrent}) can be written as
\begin{eqnarray}\nonumber
J_H\left ( T_1, T_2, H \right )&\text{=}&\int dx \mathcal{J}(x,T_1, T_2)\cos\left ( \varphi(x) \right )\text{=} \\
&\text{=}& \textup{Re} \left [ \underset{-\infty}{\overset{\infty}{\mathop \int }} dx \mathcal{J}(x,T_1, T_2) e^{i\varphi(x)} \right ],
\end{eqnarray}
where
\begin{equation}
\mathcal{J}\left (x, T_1, T_2 \right )=\int dy J_A(x,y,T_1, T_2)
\label{ThermalCurrent}
\end{equation}
is the heat current density per unit length along $x$.
The maximum value of the phase-dependent component of the thermal current is given by
\begin{equation}
J^m_H\left ( T_1, T_2,H \right )= \abs*{\int_{-\infty}^{\infty} dx \mathcal{J}(x, T_1, T_2) \cos \varphi(x)}.
\label{MaxHeatCurrent}
\end{equation}
Notice that Eq.~(\ref{MaxHeatCurrent}) resembles the expression of the maximum Josephson current $I^m_s\left ( H \right )$~\cite{Bar82}.
%\comment{Probably there is no need for additional known equations.}
%, which is given by the magnitude of the integral of $I_s\left ( H \right )$~\cite{Bar82}
%\begin{eqnarray}
%I_s( H )&\text{=}&\int dx \mathcal{I}(x)\sin \varphi(x) \text{=} \Im \left \{ \underset{\text{-} \infty}{\overset{\infty}{\mathop \int }} dx \mathcal{I}(x) e^{i\varphi(x)} \right \}
%\end{eqnarray}
%%
%that is by
%%
%\begin{equation}
%I^m_s(H)\text{=} \abs*{\int_{-\infty}^{\infty} dx \mathcal{I}(x) \cos \varphi(x)}.
%\end{equation}
%%

By assuming an uniform thermal current area density, i.e., $J_A(x,y,T_1, T_2) \equiv J_A(T_1, T_2)$, for $0\leq x \leq \nL$ and $0\leq y \leq \nW$, and zero elsewhere, Eq.~(\ref{MaxHeatCurrent}) becomes
\begin{equation}
\frac{J^m_H\left ( T_1, T_2,H \right )}{J_{int}\left ( T_1, T_2 \right )}= \frac{1}{\nL}\abs*{\int_{0}^{\nL} dx \cos \varphi(x)},
\label{MaxNormHeatCurrent}
\end{equation}
where $J_{int}\left ( T_1, T_2 \right ) = \nW \nL J_A\left ( T_1, T_2 \right )$.

It only remains to include in Eq.~(\ref{MaxNormHeatCurrent}) the proper phase difference $\varphi(x,t)$ for a long JJ.
The electrodynamics of a damped long junction is completely described in terms of a partial differential equation for the phase difference $\varphi(x,t)$, the \emph{perturbed sine-Gordon} (SG) equation~\cite{Bar82,Lik86,Val14}, 
\begin{equation}
\frac{\partial^2 \varphi }{\partial t^2}+\alpha\frac{\partial \varphi }{\partial t}-\frac{\partial^2 \varphi }{\partial x^2} = - \sin(\varphi)
\label{SGeq}
\end{equation}
where we have expressed the space variable \textit{x} in units of $\lJ$ and the time $t$ in units of the inverse of the plasma frequency $\omega_p=\sqrt{2 \pi I_c/(\Phi_0 C)}$ ($C$ is the capacitance of the junction). 
The damping parameter $\alpha=(\omega_p R_NC)^{-1}$ measures the intensity of a dissipative quasiparticle tunneling through the JJ.

 The effect of an external magnetic field $H_{ext}$ is taken into account by the boundary conditions of Eq.~(\ref{SGeq})
\begin{equation}
\frac{d\varphi(0,t) }{dx} = \frac{d\varphi(L,t) }{dx}= H,
\label{bcSGeq}
\end{equation}
where $H=2\pi\mu_0t_d\lJ/\Phi_0 H_{ext}$ is the normalized external field and $L=\nL/\lJ$ is the normalized junction length. Eqs.~\eqref{SGeq} and \eqref{bcSGeq} describe an electrically open system, namely with zero bias current and neglecting RC-loads at the ends of the junction~\cite{Sor96,Pan07}. 

The SG equation admits traveling wave solutions, called \emph{solitons}~\cite{Ust98}. 
For the unperturbed SG equation, i.e., $\alpha=0$ in Eq.~(\ref{SGeq}), solitons have the simple analytical expression~\cite{Bar82}
\begin{equation}
\varphi(x-ut)=4\arctan \left \{ \exp \left [ \pm \frac{\left(x-ut \right )}{\sqrt{1-u^2}} \right ] \right \},
\label{SGkink}
\end{equation}
where the sign $\pm$ is the polarity of the soliton and $u$ is the Swihart’s velocity, namely the largest group propagation velocity of the linear electromagnetic waves (plasma waves) in long junctions. 
A SG soliton corresponds to a phase solution changing from 0 to 2$\pi$ along the junction and has a well defined physical meaning in the long JJ framework, since it corresponds to a flux quantum $\Phi_0$ in the junction~\cite{McL82}. Thus, a soliton is usually referred to as a \emph{fluxon} or Josephson vortex in the context of long JJ. A fluxon has a width in the order of the Josephson penetration depth $\lJ$ and corresponds to a flowing supercurrent circulating around it. In this work, apart from the moment in which the external magnetic field is swept, we can essentially consider configurations of quasistatic solitons along the junction.

The number of solitons $N$ in the junction can be evaluated by the quantity~\cite{Kup06}
\begin{equation}
N=\left \lfloor \frac{\varphi(L,t)-\varphi(0,t)}{2\pi} \right\rfloor,
\label{SolitonNumber}
\end{equation}
where $\left \lfloor ... \right \rfloor$ in Eq.~(\ref{SolitonNumber}) stands for the integer part of the argument. The \emph{Meissner state} is the fluxon-free state in the junction, corresponding to $N = 0$. In the low field limit, the Meissner state $\varphi_{_M}$ and the corresponding local field $h_{_M}$ read, respectively,
\begin{eqnarray}\label{PhiMeissner}
\varphi_{_M}(x)&\sim&\frac{H_{ext}}{\cosh L/2}\sinh x \\
h_{_M}(x)&\sim&\frac{H_{ext}}{\cosh L/2}\cosh x . 
\label{LocalFieldMeissner}
\end{eqnarray}
\begin{figure*}[htbp!!]
\centering
\includegraphics[width=0.51\textwidth]{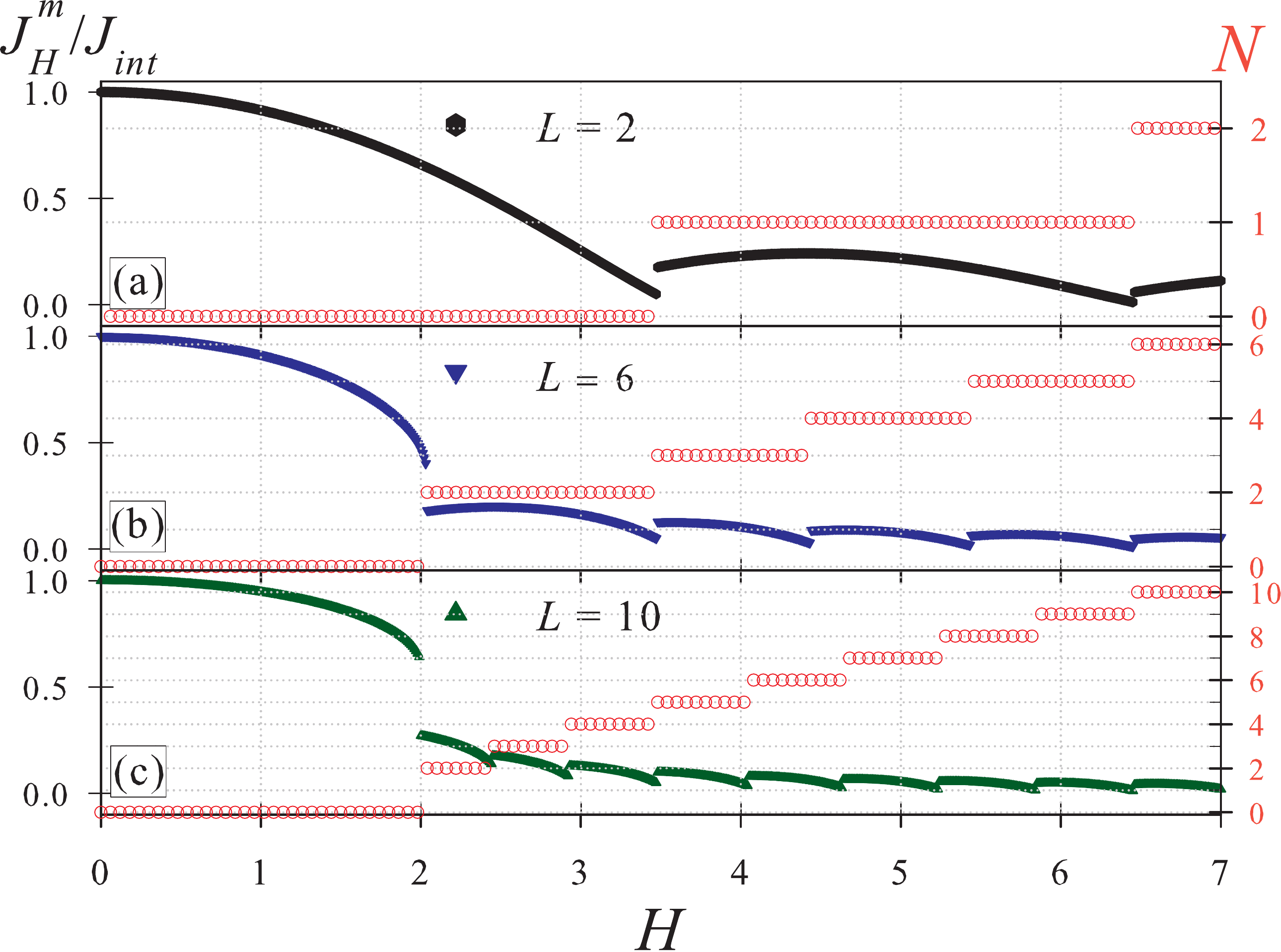}
\includegraphics[width=0.47\textwidth]{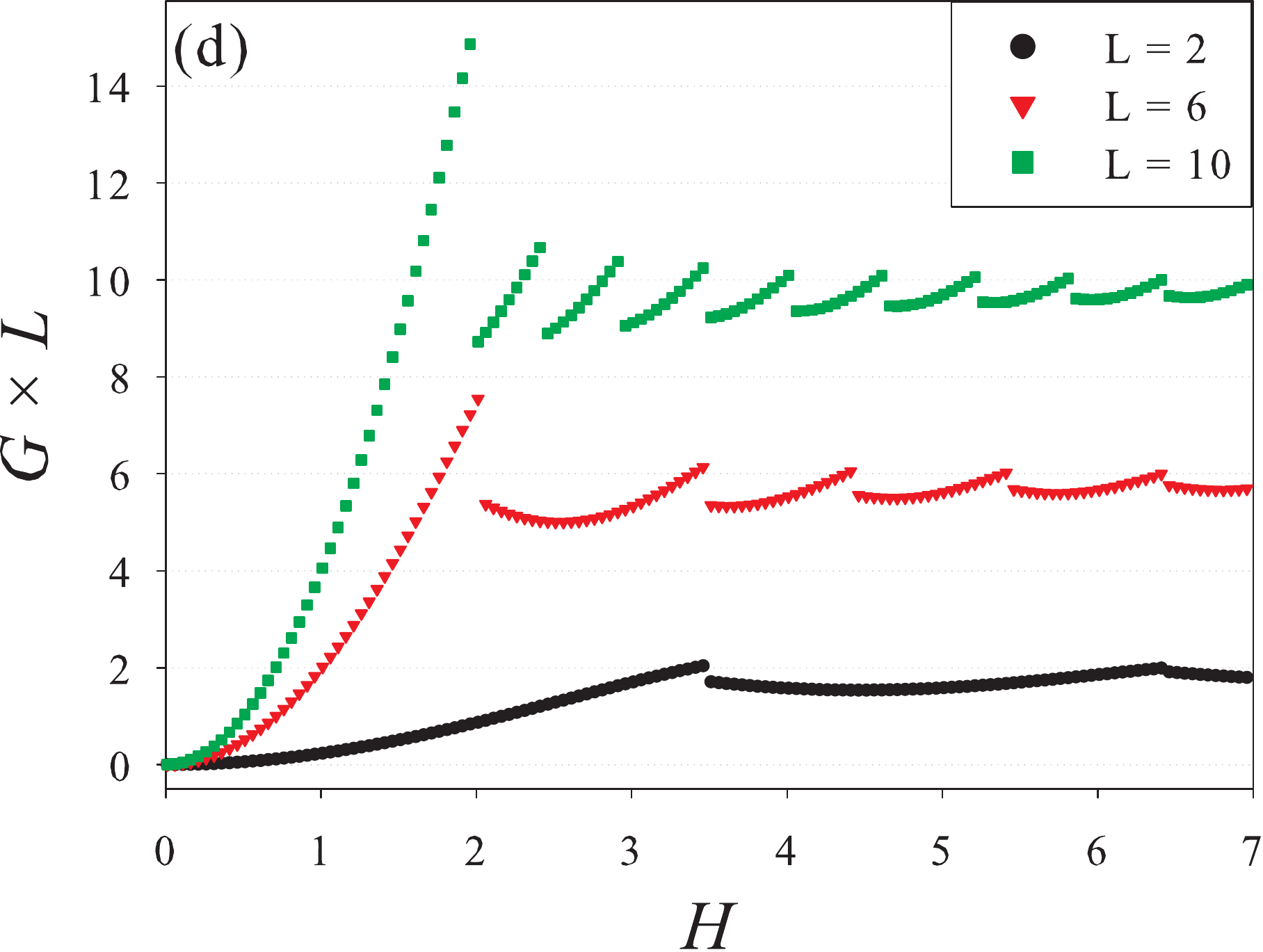}
\caption{(Color online) Normalized maximum heat current $J^m_H/J_{int}$ (left ordinate scale, full symbols) and soliton number $N$ (right ordinate scale, empty symbols) as a function of the sweeping up magnetic field $H$ for JJ lengths $L=2, 6,10$ and damping parameter $\alpha = 0.1$ [left panels (a), (b), and (c)]. In panel (d), Gibbs free energy $G$ [see Eqs.~(\ref{E})] as a function of the sweeping up magnetic field $H$, for several JJ lengths $L=2, 6,10$ and damping parameter $\alpha=0.1$. The multiplication per $L$ is used as an offset to avoid the superimposition of the curves.
} 
\label{Fig2}
\end{figure*}

Several authors~\cite{Yug95,Yug99,Kup06,Kup10} faced the study of SG solutions by analyzing the Gibbs free-energy functional and its minimization (at least locally).
The Gibbs free energy $G$ of a long JJ consists of the Josephson coupling energy $E_J$, and the magnetic and electrical energies $E_m$ and $E_e$, respectively given by~\cite{Lik86, Cas96}
\begin{eqnarray}
E_J&=&\frac{E_{J_0}}{L}\int_0^Ldx\left (1-\cos\varphi\right)\\
\label{EJ}
E_m&=&\frac{E_{J_0}}{L}\int_0^Ldx\frac{1}{2}\left ( \frac{d\varphi}{dx}-H \right )^2 \\
\label{Em}
E_e&=&\frac{E_{J_0}}{L}\int_0^Ldx \frac{1}{2} \left ( \frac{d\varphi}{dt}\right )^2, 
\label{Ee}
\end{eqnarray}
where $E_{J_0}=\Phi_0 I_c/2\pi$. The magnetic energy $E_m$ includes both the energy stored in the total inductance of the JJ and the energy supplied by the magnetic field. 
%According to the slow evolution of $\varphi$, we can rightly ignore the electric term, so that the Gibbs free energy, normalized to $E_{J_0}$, reads
In case of static or slowly changing magnetic field, the $E_e$ contribution in Eq. (\ref{Ee}) can be neglected and the Gibbs free energy, normalized to $E_{J_0}$, reads
\begin{eqnarray}\label{E}
G&=&\frac{E_J}{E_{J_0}}+\frac{E_m}{E_{J_0}}=\\ \nonumber
&=&\frac{1}{L}\int_0^Ldx\left \{ \left (1-\cos\varphi\right )+\frac{1}{2}\left ( \frac{d\varphi}{dx}-H \right )^2 \right \}.
\end{eqnarray}
This expression can be minimized to find which phase solution is the most stable one ~\cite{Yug95,Yug99,Kup06,Kup10}.

However, as we will show below, this static and thermodynamics approach is in general not sufficient because it fails to describe the JJ state when multiple solutions are available.
Here we rely on a full dynamical description and solve Eqs.~(\ref{SGeq}) and (\ref{bcSGeq}) in a ``quasistationary'' approach with slowly-varying boundary conditions. Specifically, the magnetic field $H(t)$ is modeled as a staircase function with small steps $\Delta H=0.01$ kept constant for time intervals long enough to ensure the restoring of a steady phase configuration before the field $H$ is further modified. This sweeping method closely maps an experimental setup used to obtain diffraction patterns.
The analysis is performed by numerical integration of the system~(\ref{SGeq}) and (\ref{bcSGeq}) using flat initial conditions, $\varphi(x,0)=d\varphi(x,0)/dt=0\quad \forall x\in[0-L]$, and time and spatial integration steps fixed at $\Delta t = 0.01$ and $\Delta x = 0.01$, respectively.

\section{The Results}
\label{Results}

\subsection{Forward dynamics}
\label{ResultsA}

The analysis is carried out studying, according to Eqs.~(\ref{MaxNormHeatCurrent}), (\ref{SGeq}) and (\ref{bcSGeq}), the normalized maximum heat current $J^m_H/J_{int}$ as a function of the magnetic field $H$, by varying the junction length $L$. Moreover, different dissipative regimes are taken into account by changing the damping parameter $\alpha$, in order to range from underdamped ($\alpha=0.1$) to overdamped ($\alpha=10$) conditions.

%\comment{Let us first focus on the magnetic field dependence for a fixed length, e.g., $L=2$ in Fig. \ref{Fig2} (a).
Let us first focus on the magnetic field dependence for fixed lengths, e.g., $L=2, 6, 10$, setting $\alpha=0.1$ in panels (a), (b), and (c) of Fig.~\ref{Fig2}.
The magnetic field dependence of $J^m_H$ results in ``Fraunhofer-like'' diffraction patterns.
While in the short junction limit~\cite{Gia13,Mar14}, different diffraction \emph{lobes} are well separated, here we observe the overlapping of the lobes.
The transitions between these lobes is usually discontinuous.
These pattern can be explained in terms of solitons entering the JJ. 

Each lobe corresponds to a state with a fixed number $N$ of solitons.
When the magnetic field increases, the configuration with more solitons is energetically favorable and, thus, the system jumps from a metastable state to a more stable state with more solitons.
In the region of $H$ values in which the diffraction lobes overlap, several solutions with different $N$ may concurrently exist.
Therefore, the system stays in the present configuration until the following one is energetically more stable.

Experimental evidences of such transitions from metastable to stable states in the diffraction patterns of the maximum Josephson current $I^m_s(H)$ in long~\cite{Owe67,Mat69,Cir97}, annular~\cite{Kei96,Mar96,Fra00} and grain boundary~\cite{Mit99} JJs has been observed. 
This thermodynamical and energetic view is confirmed by the study of the Gibbs free energy as a function of $H$.
The Gibbs free energy jumps to a new local minimum when new solitons penetrate into the junction, as it stands out in panel (d) of Fig.~\ref{Fig2}.

\begin{figure*}[htbp!!]
\centering
\includegraphics[width=\textwidth]{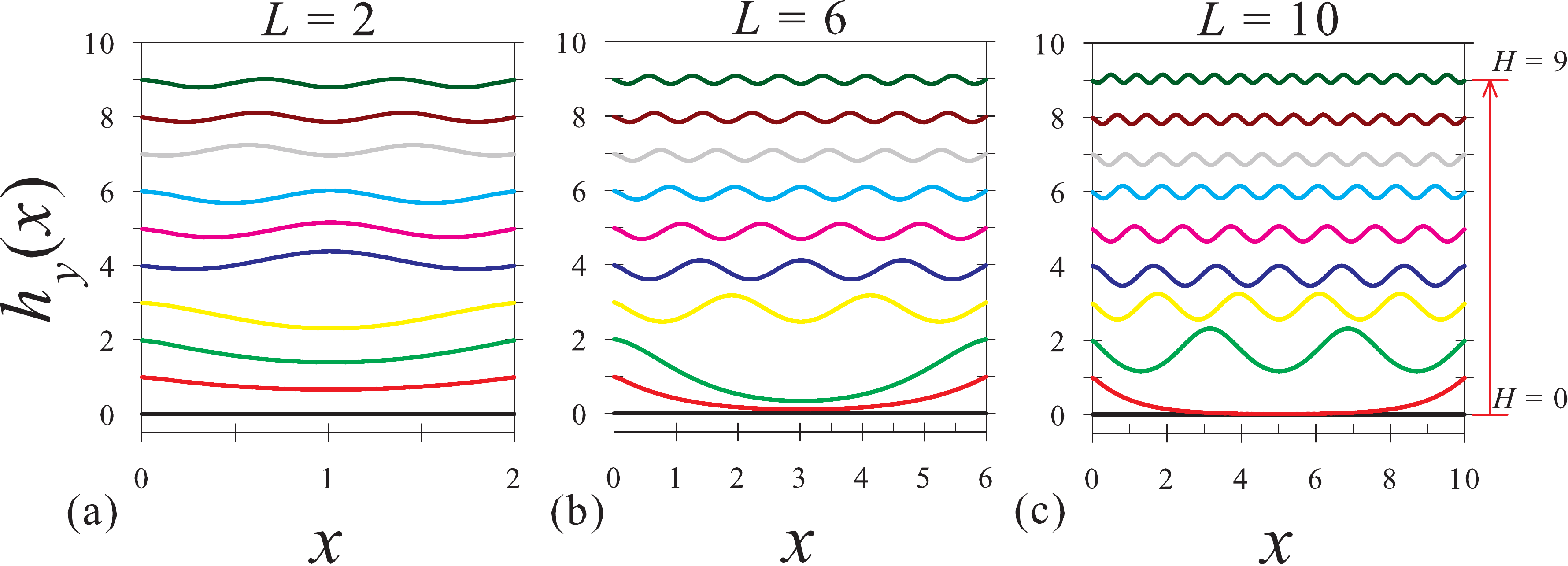}
\caption{(Color online) Local magnetic field $h_y(x)$ [see Eq.~(\ref{LocalMagneticField})] as a function of $x$, for damping parameter $\alpha=0.10$ and several JJ lengths $L = 2, 6, 10$ [panels (a), (b) and (c), respectively]. These results are obtained when increasing $H$ in steps of $\Delta H=1.0$ from $H=0$ (black line) to $H=9$ (dark-green line).
}
\label{Fig3}
\end{figure*}
\begin{figure}[b!!]
\centering
\includegraphics[width=0.49\textwidth]{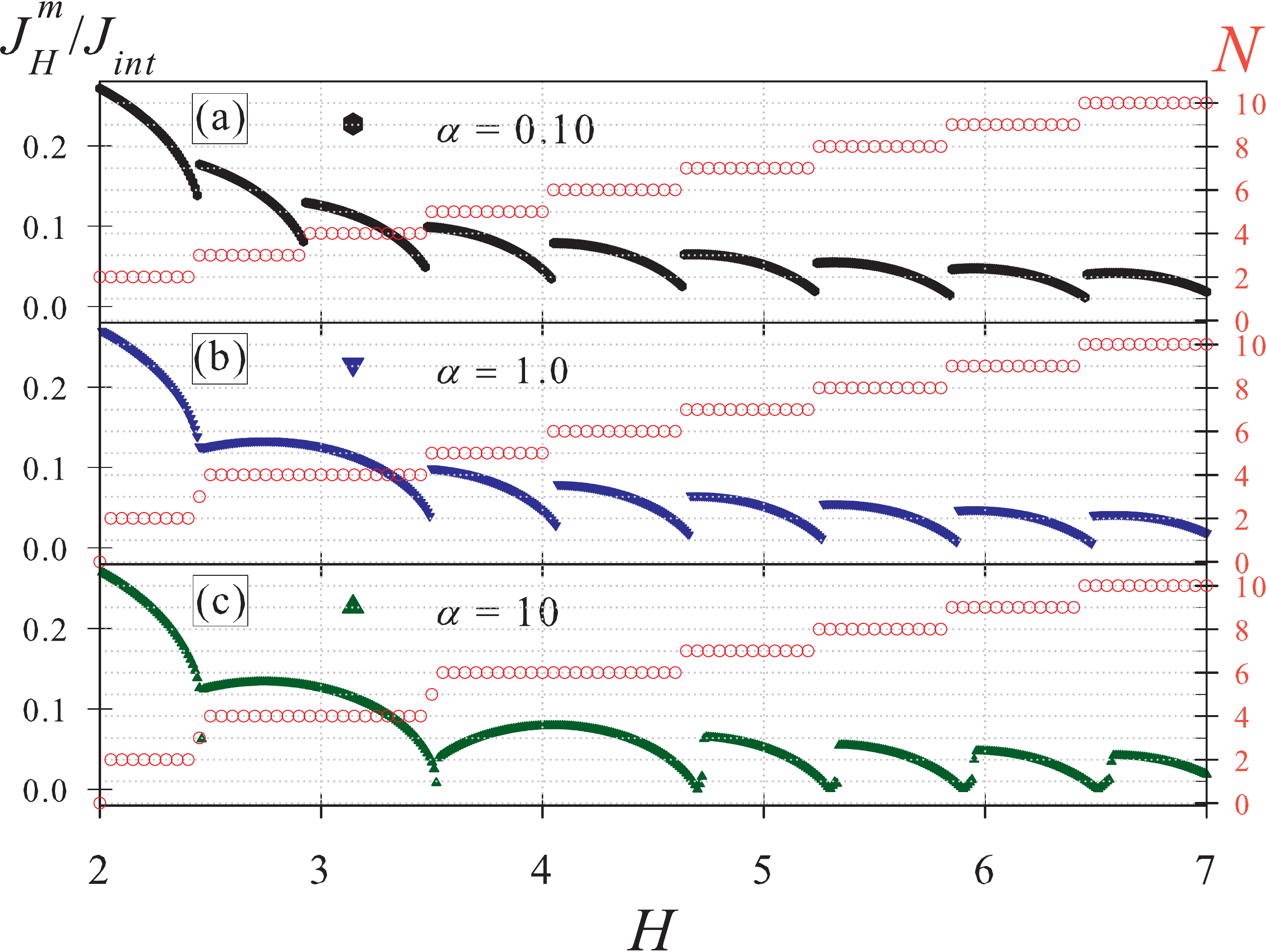}
\caption{(Color online) Normalized maximum heat current $J^m_H/J_{int}$ (left ordinate scale, full symbols) and soliton number $N$ (right ordinate scale, empty symbols) as a function of the sweeping up magnetic field $H$ for $L=10$ and $\alpha = 0.10, 1.0, 10$ [panels (a), (b), and (c), respectively]. 
}
\label{Fig4}
\end{figure}

In panels (a), (b), and (c) of Fig.~\ref{Fig2}, as long as $H$ increases up to a critical value $H_c$, the normalized heat current monotonically reduces. Specifically, $H_c\simeq 3.5$ for $L=2$, see panel (a) of Fig.~\ref{Fig2}, and $H_c=2$ for $L=6, 10$, see panels (b) and (c) of Fig.~\ref{Fig2}. By exceeding the threshold value $H_c$, the second lobe begins. The first lobe corresponds to $N=0$, i.e., Meissner states with zero solitons in the junction.
This value of the critical field characterizes also the diffraction patterns of the maximum Josephson current $I^m_s(H)$ in both overlap and inline long JJs~\cite{Owe67,Bar82,Cir97}. Above the critical field, solitons in the form of magnetic fluxons can penetrate into the junction, resulting in $N>0$. For $L=2$, see panel (a) of Fig.~\ref{Fig2}, any transition between neighboring lobes corresponds to increase $N$ by one per lobe. Conversely for longer junctions, for $H$ just above $H_c$ the system switches from the Meissner to a two-solitons state, in which a pair of fluxons is symmetrically injected from the junction edges. For high fields, $N$ advances again by one per lobe, as shown in panels (b) and (c) of Fig.~\ref{Fig2}. 

A complete description of the junction dynamics is presented in Fig.~\ref{Fig3}.
The spatial distributions of the local magnetic field $h_y(x)$, calculated according to Eq.~(\ref{LocalMagneticField}), are shown in Fig.~\ref{Fig3}, for $L=2,6,10$ and $\alpha=0.1$. The ripples in these curves indicate fluxons along the junction. For $H<H_c$ the system is in the Meissner state, see Eq.~(\ref{LocalFieldMeissner}), i.e., no ripples, meaning zero fluxons, and a decaying magnetic field penetrating the junction ends. According to the nonlinearity of the problem, for high fields ($H>H_c$) the stable solutions are not the trivial superimposition of Meissner and vortex fields, but are rather solitons ``dressed'' by a Meissner field confined in the junction edges~\cite{Kup06}. The amount of fluxons, e.g., ripples, along the JJ increases by intensifying the magnetic field. Moreover, for a fixed value of $H$, the number of fluxons grows as the junction length increases, since a longer junction means more space to arrange ``ripples'' along it.

\begin{figure*}[htbp!!]
\centering
\includegraphics[width=\textwidth]{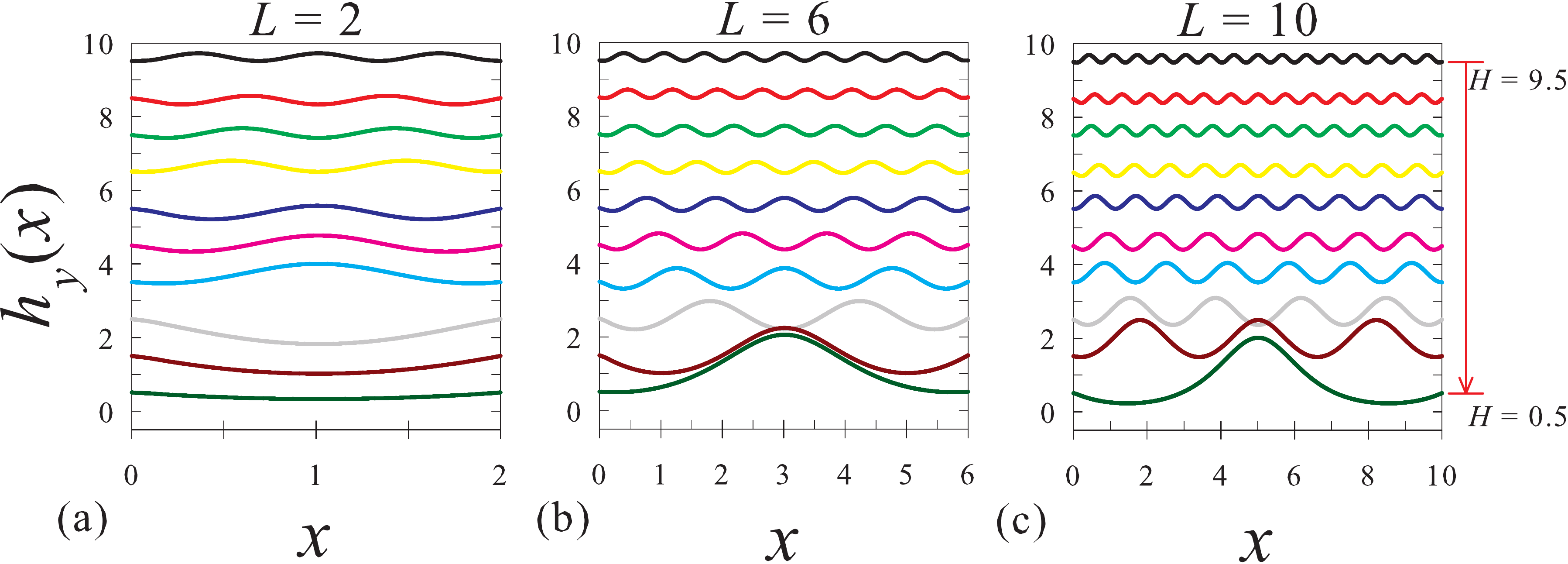}
\caption{(Color online) Local magnetic field $h_y(x)$ [see Eq.~(\ref{LocalMagneticField})] as a function of $x$, for damping parameter $\alpha=0.10$ and several JJ lengths $L = 2, 6, 10$ [panels (a), (b) and (c), respectively]. These results are calculated as the value of $H$ is reduced in steps of $\Delta H=1.0$ from $H=9.5$ (black line) to $H=0.5$ (dark-green line), after an initial increasing sweep
}
\label{Fig5}
\end{figure*}

Up to now we have interpreted the diffraction patterns in terms of transition between a metastable and stable state.
This thermodynamical picture in not complete, as the numerical calculation for different damping parameter $\alpha$ shows (see Fig.~\ref{Fig4}).
As $\alpha$ increases the diffraction pattern changes and some of the lobes vanish.
For instance, for $L=10$ as $\alpha$ increases from 0.1 to 1.0, the lobe of $J^m_H(H)$ for $N=3$ vanishes being replaced by a large one corresponding to four-solitons solutions [see panels (a) and (b) of Fig.~\ref{Fig4}]. 
Similarly, the lobe for $N=5$ for $\alpha=1.0$ [panel (b) of Fig.~\ref{Fig4}] is replaced by a large lobe describing configurations with $N=6$ solitons for $\alpha=10$ [panel (c) of Fig.~\ref{Fig4}]. 
Since $\alpha$ is a damping parameter, these results cannot be explained in terms of Gibbs free energy minimization but we need a dynamics interpretation according to Eq.~(\ref{SGeq}).\\
\indent From a qualitative point of view, we relate the entering of solitons to $2\pi-$jumps in the phase [see Eq.~(\ref{SolitonNumber})].
In a static thermodynamics view, this is a sudden transition between metastable and stable solutions.
However, Eq.~(\ref{SGeq}) gives us the full dynamics of the transitions which is influenced by $\alpha$.\\
\indent As a friction parameter, $\alpha$ opposes the variations of $\varphi$ and, if it is large enough, can stabilize a state otherwise metastable.
Consequently, even if for a given $H$ there are solutions that are energetically favorable, the system is dynamically frozen in a metastable one.
Although this \textit{per se} explains why by increasing $\alpha$ some lobes vanish until new solitons can enter the junction, we can go beyond by looking at the solitons arrangement along the junction when the magnetic field is swept.\\
\indent For the sake of symmetry, we first observe that in the midpoint of the system, i.e., $x=L/2$, the local field $\eta_i$ can be exclusively a local maximum (accordingly, a soliton is pinned in the center of the junction and the total amount of solitons is odd) or a local minimum (the total amount of solitons is even) (see curves in Figs.~\ref{Fig3} and~\ref{Fig5} for $x=L/2$). As the local field at the midpoint, i.e., $h_y(x\text{=}L/2)$, sudden switches from a minimum to a maximum (or vice versa), the diffraction patterns abruptly jumps and, accordingly, the soliton number $N$ suddenly changes by one. \\
\indent Due to the symmetry of the system, as the intensity of the magnetic field increases, the solitons set along the junction are shifted towards its center. For low fields, this shift is strongly affected by the damping, so that solitons are less and less pushed by the external field as the damping increases. Let's suppose that the junction contains an even number of solitons. When the external field value is such that a new pair of solitons is injected, if the two solitons near the center are close enough they can merge in a midpoint soliton. Correspondingly, the total amount $N$ of solitons gets odd. However, high damping can prevent this phenomenon, so that, as a result of the solitons injection, $N$ changes by two and the total amount of solitons remains even.

\subsection{Forward and backward dynamics and hysteretic effect}
\label{ResultsB}

We have seen that the dissipative dynamics is crucial to understand the junction diffraction patterns.
In general, the dissipative dynamics depends on the full evolution of the system.
Therefore, it is natural to wonder if the heat diffraction pattern changes with the history of the system.

\begin{figure*}[htbp!!]
\centering
\includegraphics[width=0.506\textwidth]{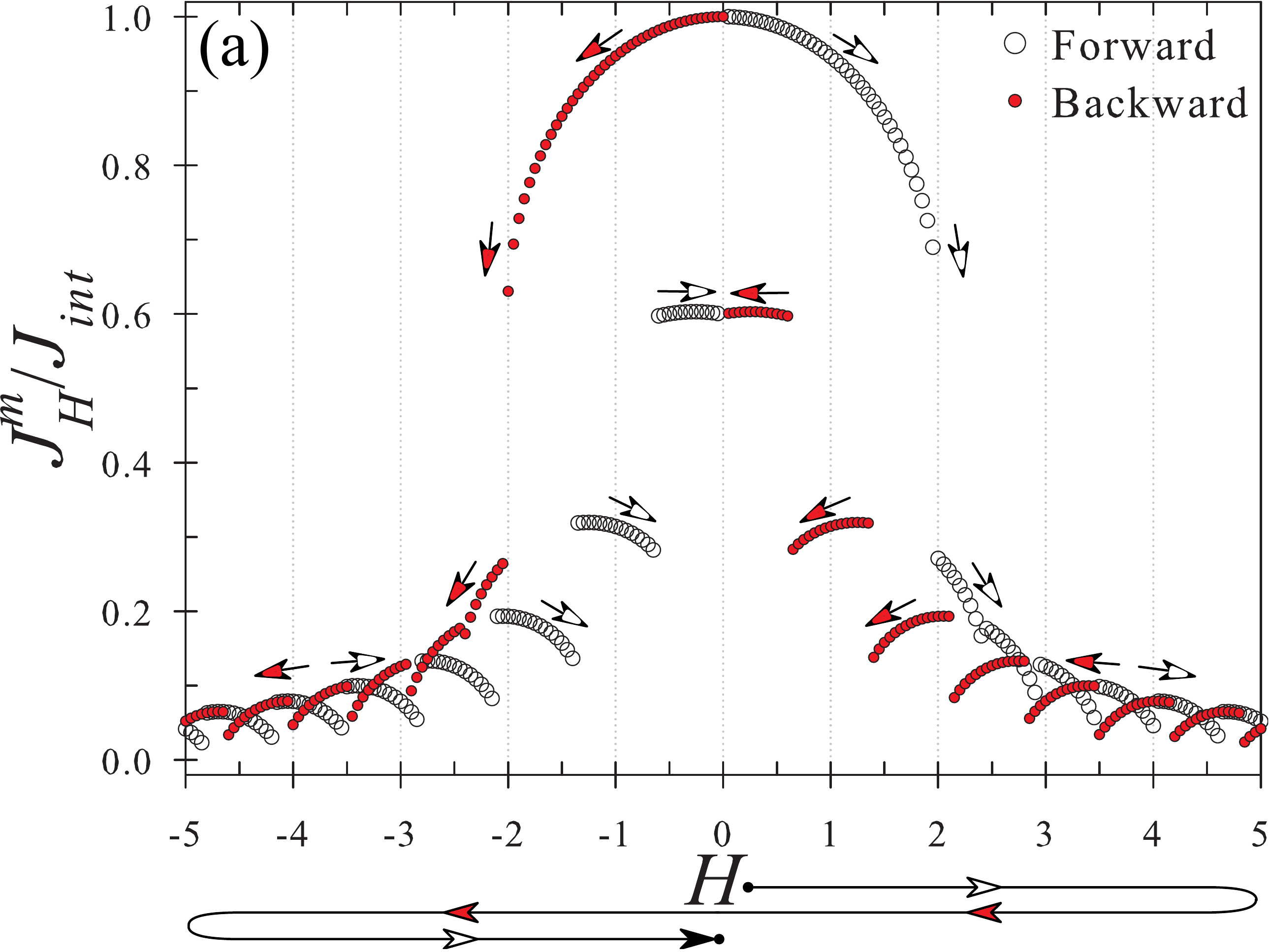}
\includegraphics[width=0.484\textwidth]{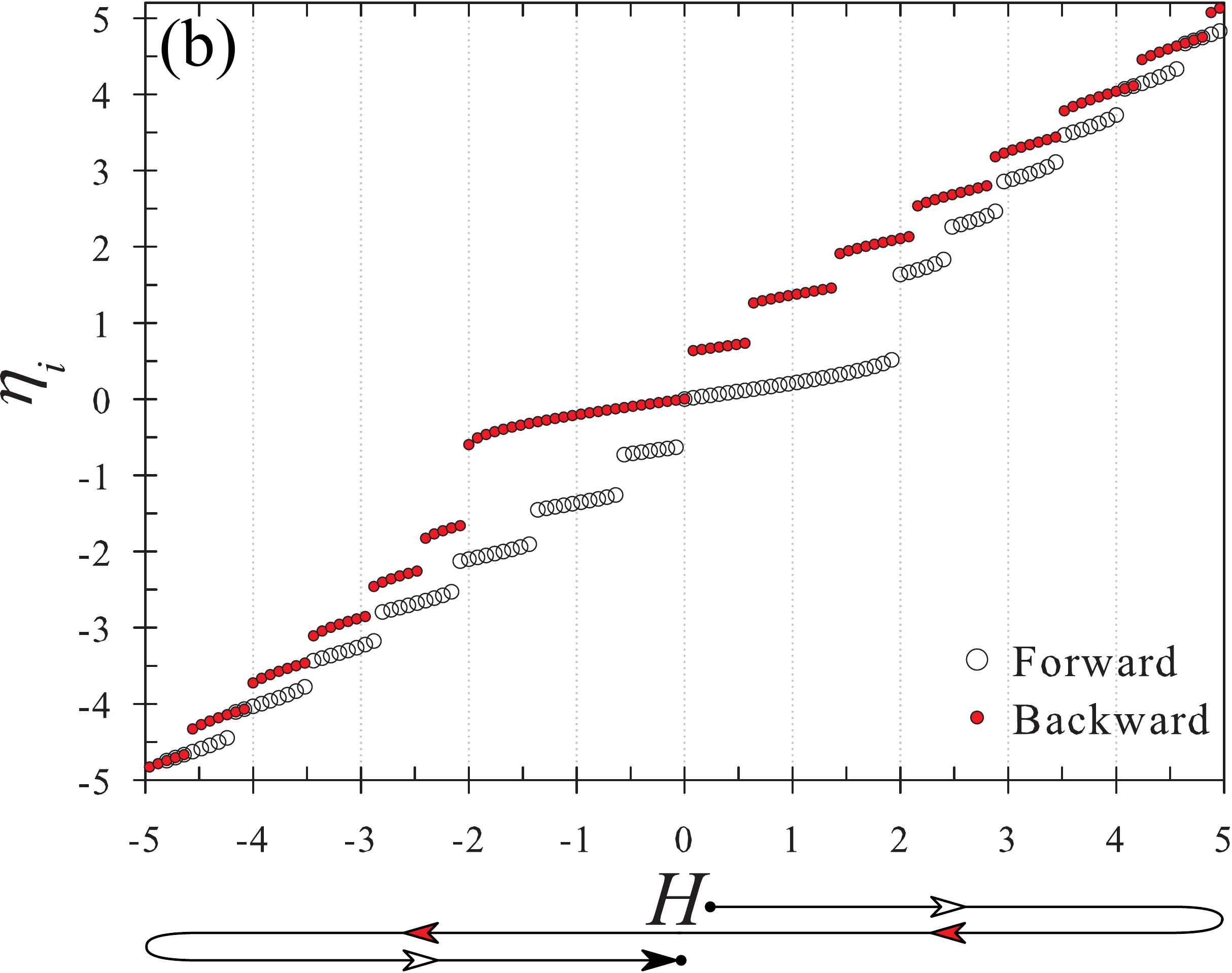}
\caption{(Color online) The normalized maximum heat current $J^m_H/J_{int}$ [panel (a)] and the mean local field $\eta_i$, i.e. the magnetic flux through the junction per unit area, [panel (b)], setting the length and the damping parameter to $L=10$ and $\alpha = 0.1$, respectively, as a function of the magnetic field $H$. Specifically, $H$ is swept forward (empty circle) from $H\text{=}0$ to $H\text{=}5$, then backward (full red circle) from $H\text{=}5$ to $H\text{=}-5$ and again forward (empty circle) from $H\text{=}-5$ to $H\text{=}0$. For clarity, in panel (a) forward and backward directions of $H$ are marked with empty and full red arrows, respectively. }
\label{Fig6}
\end{figure*}

To answer this question, we have implemented a double-swept drive.
The external ``staircase'' field $H$ is ramped up from $H=0$ to $H_{max}$, then reduced to $-H_{max}$ and subsequently raised again to zero.
The corresponding heat diffraction pattern is shown in Fig.~\ref{Fig6} (a), for $L=10$ and $\alpha=0.1$.
As we can see, the forward, i.e., with $H$ increasing, and the backward, i.e., with $H$ decreasing, patterns are significantly different.
For a given value of the magnetic field, the heat transferred in the backward and forward evolutions are usually different and the system is found in a different diffraction lobe.
Following Fig.~\ref{Fig2}, we can associate the forward and backward stable states (at fixed $H$) with a different number of solitons in the junction. 

The explanation of this difference can be found again in the damped dynamics. 
Suppose that, for a given $H$, we have $\tilde{N}$ solitons in the junction, so that new solitons are injected into (extracted from) the system by increasing (decreasing) the magnetic field.
Let's also assume that injection of solitons occurs for $H=\tilde{H}$, so that now $\tilde{N}+1$ solitons set in the junction.
In the backward evolution, when $H$ reaches the value $\tilde{H}$, the dissipation terms opposes to the phase variations, so that $(\tilde{N}+1)$-solitons solutions can still be the most favorable ones.
Therefore, the system remains in the $\tilde{N}+1$ lobe even if $\tilde{H}$ is passed. 
This effect generates the asymmetry between the forward and backward evolution.

Interestingly, the overall effect is a hysteric behavior in the heat power [see Fig.~\ref{Fig6}(a)].
The heat power difference $(J_H^{for}-J_H^{back})/J_{int}$ can be as large as $0.65$ for $H\simeq 1$ [see Fig.~\ref{Fig6}(a)], $J_H^{for}$ and $J_H^{back}$ being the forward and backward maximum heat current densities, respectively.
This strong \emph{heat hysteresis} paves the way to interesting implementations of the extended long JJs.
For instance, it could be used as a heat memory by storing a ``heat bit $1$'' if $J_H^{for}/J_{int}>0.9$ and ``a heat bit $0$'' if $J_H^{back}/J_{int}<0.4$ at $H\simeq1$. Specifically, we are suggesting a memory element in which the I/O related variables defining the history-dependent behavior of the device~\cite{Per11} are the external magnetic field $H(t)$ and the maximum heat currents $J_H^{for}(t)$ and $J_H^{back}(t)$, respectively. The memory states could be represented by the distinct diffraction lobes in that ranges of magnetic field, such as $H\in[0-2]$, in which the forward/backward diffraction patterns clearly differ.

Hysteresis results also studying the local magnetic field $h_y(x)$ for a backward sweeping magnetic field [see Fig.~\ref{Fig5}]. Specifically, for sufficiently long JJs, see panels (b) and (c) of Fig.~\ref{Fig5} for $L=6$ and 10, respectively, when switching off the field, i.e., bottom curves for $H=0.5$, a fluxon remains confined in the midpoint of the junction. 
 
The hysteretic behavior as a function of the magnetic field clearly manifests also in the magnetic flux $\Phi_i$ through the junction per unit area. This quantity coincides with the mean value $\eta_i$ of the local field according to
\begin{equation}
\frac{\Phi_i}{A}=\frac{1}{t_d L}\underset{0}{\overset{t_d}{\mathop \int }} dy \underset{0}{\overset{L}{\mathop \int }} h_y(x) dx=\frac{1}{L}\underset{0}{\overset{L}{\mathop \int }} h_y(x) dx=\eta_i,
\label{MagneticFlux}
\end{equation}
$A=t_d L$ being the effective magnetic area of the junction.
Panel (b) of Fig.~\ref{Fig6} shows the mean local field $\eta_i$ as a function of $H$, for $L=10$ and $\alpha = 0.1$. We observe that $\eta_i\lesssim H$ ($\eta_i\gtrsim H$) when sweeping forward (backward) the magnetic field. Each branch of these curves corresponds to configurations with a specific amount of fluxons. For $|H|\leq H_c$, we note a long branch around $\eta_i\sim 0$ corresponding to the Meissner field and three branches with higher values of $|\eta_i|$ corresponding to configurations with $N=1,2,3$ fluxons. For high field intensities, both forward and backward $\eta_i$ curves approach the $H$ values, meaning full penetration of strong external field inside the junction.

\subsection{Effect of the temperature}
\label{ResultsC}

We take also into account the effect of the temperature on the phase dynamics by including the thermal fluctuations in the SG model~\cite{Val14,Gua16}
\begin{equation}
\frac{\partial^2 \varphi }{\partial t^2}+\alpha\frac{\partial \varphi }{\partial t}-\frac{\partial^2 \varphi }{\partial x^2} = - \sin(\varphi)+i_f(x,t)
\label{noisySGeq}
\end{equation}
with boundary conditions still given by Eq.~(\ref{bcSGeq}).
The normalized thermal current $i_f\left ({x,t}\right )$ is characterized by the well-known statistical properties of a Gaussian random process
\begin{figure*}[htbp!!]
\centering
\includegraphics[width=\textwidth]{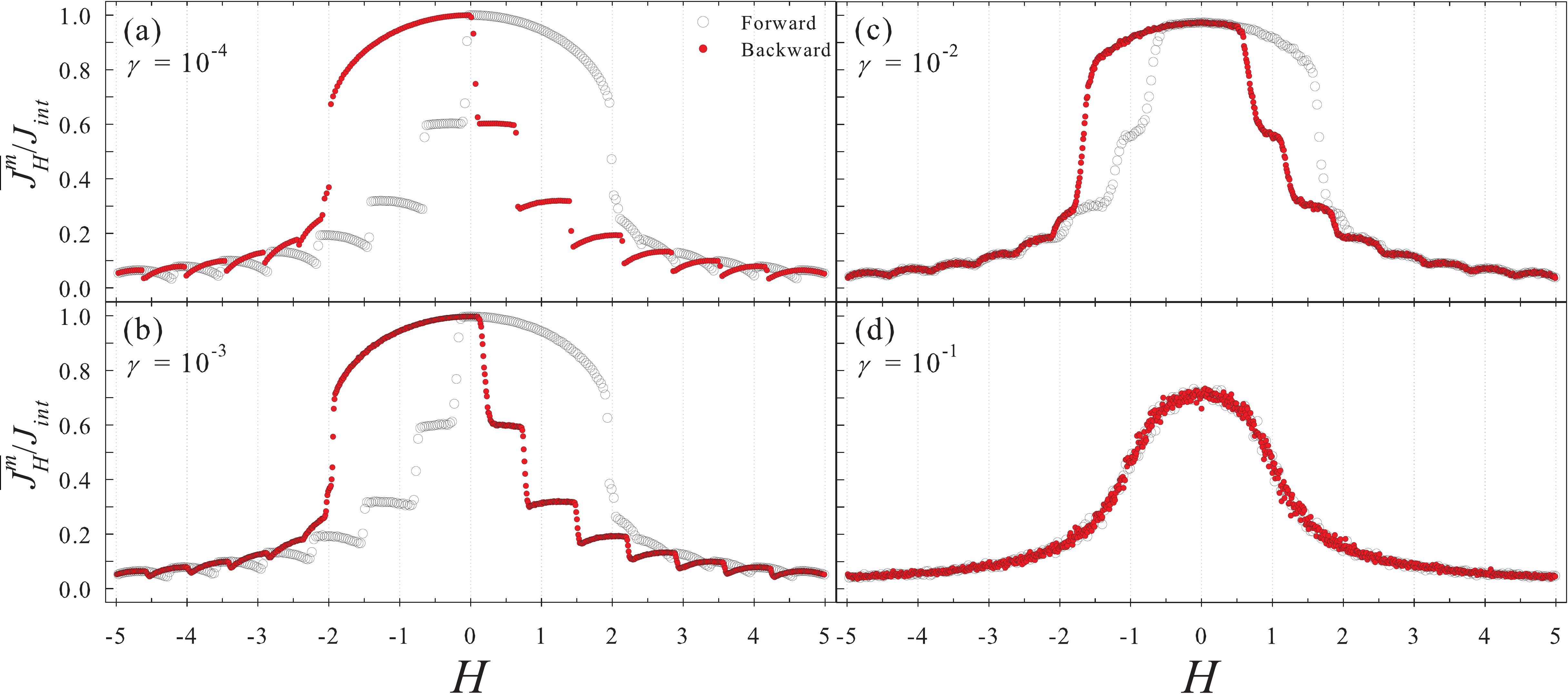}
\caption{(Color online) Average value of the normalized maximum heat current $\overline{J^m_H}/J_{int}$ over $N_{exp}=100$ numerical realizations as a function of the magnetic field $H$, setting the length and the damping parameter to $L=10$ and $\alpha = 0.1$, respectively, for different values of the noise intensity $\gamma =$ $10^{-4}$ (a), $10^{-3}$ (b), $10^{-2}$ (c), and $10^{-1}$ (d). Specifically, $H$ is swept forward (empty circle) from $H\text{=}0$ to $H\text{=}5$, then backward (full red circle) from $H\text{=}5$ to $H\text{=}-5$ and again forward (empty circle) from $H\text{=}-5$ to $H\text{=}0$. The legend in panel (a) refers to all panels.} 
\label{Fig7}
\end{figure*}
\begin{eqnarray}
\left \langle i_f \left ({x,t}\right ) \right \rangle &=& 0, \\
 \left \langle i_f\left ({x,t}\right )i_f\left ({x',t'}\right ) \right \rangle &=& 2\gamma(T) \delta \left (x-x' \right )\delta \left (t-t' \right ),
\label{WNProperties}
\end{eqnarray}
where $\delta$ is the Dirac delta function and the noise intensity $\gamma$ is proportional to the temperature $T$ according to~\cite{Cas96}
\begin{equation}\label{WNAmp}
\gamma(T)=\frac{2\pi }{\Phi_0}L\alpha\frac{k_bT}{I_c(T)}.
\end{equation}
For instance, for an Nb/AlO$_x$/Nb long JJ with a critical current $I_c(0)=2.7\textup{mA}$ and $L=10$, the noise amplitudes $\gamma = 10^{-4}, 10^{-3},10^{-2}, 10^{-1}$ correspond to the temperatures $T = 0.034, 0.34, 3.3, 8.1 \textup{K}$, respectively.\\
\indent With the aim of observing the overall effect of the thermal fluctuations on the diffraction patterns, we can in first approximation consider the temperature $T$ in Eq.~(\ref{WNAmp}) as the temperature $T_1$ of the hot electrode.\\
\indent According to the stochastic nature of Eq.~(\ref{noisySGeq}), the quantity $\overline{J^m_H}/J_{int}$ is computed by averaging the normalized maximum heat current over the total number of numerical realizations $N_{exp}=100$. Results for $L=10$, in underdamped conditions $\alpha = 0.1$, and with different values of the noise intensity $\gamma = 10^{-4}, 10^{-3}, 10^{-2}$, and $10^{-1}$, are shown in panels (a), (b), (c), and (d) of Fig.~\ref{Fig7}, respectively. By increasing the noise intensity, we obtain a smoothing of the diffraction patterns with broadened transitions between stable configurations. Also the hysteretic behavior is affected by an enhancement in the noise intensity. Moreover, as the noise increases, the ``Fraunhofer-like'' structure of the patterns persists up to $\gamma$ values of the order of the activation energy $\Delta U$ of the thermally induced phase slippage in long JJs~\cite{Cas96}. In fact, for $\gamma=0.1$, forward and backward $J^m_H(H)$ curves are superimposed, resulting in a large peak centered in $H=0$, see panel (d) of Fig.~\ref{Fig7}.\\
\indent The effect of the temperature has to be also taken into account for a proper normalization of the junction length $\nL$. In fact, according to Eq.~(\ref{LambdaJ}), the Josephson penetration length $\lJ$ depends on the temperatures $T_1$ and $T_2$ of the electrodes $S_1$ and $S_2$ through both the critical current $I_c(T_1,T_2)$ and the effective magnetic thickness $t_d(T_1,T_2)$. Let us suppose that the electrodes are made by the same superconductors, so that $\lambda_i(0)\equiv\lambda(0)$, $T^i_c\equiv T_c$, and $\Delta_i \left ( T_i \right )\equiv\Delta \left ( T_i \right )$.\\
\indent For a temperature biased JJ the critical current $I_c( T_1,T_2)$ reads~\cite{Gia05,Tir08,Bos16}
\begin{eqnarray}\label{IcT1T2}\nonumber
I_c\left ( T_1,T_2 \right )=&&\frac{1}{2eR_N}\Big |\underset{-\infty}{\overset{\infty}{\mathop \int }} d\varepsilon \{ f\left ( \varepsilon ,T_1 \right )\textup{Re}\left [\mathfrak{F}_{S_1}(\varepsilon ) \right ]\textup{Im}\left [\mathfrak{F}_{S_2}(\varepsilon ) \right ] \\
&&+ f\left ( \varepsilon ,T_2 \right )\textup{Re}\left [\mathfrak{F}_{S_2}(\varepsilon ) \right ]\textup{Im}\left [\mathfrak{F}_{S_1}(\varepsilon ) \right ] \} \Big |
\end{eqnarray}
where $\mathfrak{F}_{S_j}(\varepsilon ) =\Delta \left ( T_j \right )\Big/\sqrt{\left ( \varepsilon +i\Gamma_j \right )^2-\Delta^2 \left ( T_j\right )}$, $\Gamma_j$ being the Dynes parameter~\cite{Dyn78}. The temperature-dependent Josephson length can be written as
\begin{equation}\label{LambdaJT1T2}
\frac{\lambda_J\left ( T_1,T_2 \right )}{\lambda_J\left ( 0,0 \right )}=\sqrt{\frac{I_c(0,0)t_d(0,0)}{I_c(T_1,T_2)t_d(T_1,T_2)}}.
\end{equation}
\indent By increasing the temperatures $T_1$ and $T_2$, also $\lJ( T_1,T_2)$ slightly increases. Specifically, $\lJ( T_1,T_2)/\lJ( 0,0)\sim2$ for $T_i\rightarrow T_c$, with $i=1,2$ (setting $\Gamma_i/\Delta(0)\text{=}10^{-7}$)~\cite{Sai12}. Thus a JJ which is long, i.e., $L(T_1,T_2)\text{=}\nL/\lJ(T_1,T_2)\text{>}1$, at low temperatures becomes smaller and smaller when $T_i$ approach $T_c$. Nevertheless, for $L(T_1,T_2)\text{>}2$ the ``long junction'' condition on the JJ length is always satisfied, regardless of the temperature $T_i$ of the superconductors. Since $\lJ(T_1,T_2)\text{>}\lJ(0,0)$, also the condition on the JJ width is clearly satisfied, i.e., $W(T_1,T_2)\text{=}\nW/ \lJ(T_1,T_2)\ll1$.

\section{Conclusions}
\label{Conclusions}

We have analyzed the diffractions patterns of the heat current in a thermally-biased long Josephson junction (JJ) when an external magnetic field $H$ is properly swept. In particular, we studied the maximum heat current as a function of $H$ by varying the length and the damping of the junction. The phase dynamics is analyzed within the sine-Gordon (SG) framework with proper boundary conditions modeling a slowly-changing external magnetic field. We have shown the lobes structure of the heat current diffraction patterns, which is closely related to the Josephson vortices, i.e., solitons, injections through the edges and arrangement along the JJ length. The amount of solitons set along the system depends mainly on the junction length and the magnetic field intensity, but we observed modifications in the diffraction patterns when the value of the parameter $\alpha$, describing the damping of the system, is increased. This phenomenon can be understood by leaving the pure thermodynamical picture of the problem, and exploring the full dynamics of the SG model. Moreover, the study of the full evolution of the system disclosed a clear hysteretic effect as a function of $H$ in both the diffraction pattern and the magnetic flux per unit area through the junction. Finally, we analyzed noise-induced effects in the diffraction patterns by including a Gaussian thermal noise source into the SG model.\\
\indent The proposed systems could be easily implemented by standard nanofabrication techniques through the setup proposed for the short JJs-based thermal diffractor~\cite{Mar14}. Besides being relevant from a fundamental physics viewpoint, the studied thermally-biased long JJ represents the first effort to combine the physics of solitons and the emerging superconducting coherent caloritronics, e.g., the implementation of Josephson-based heat memories.\\
\begin{acknowledgments}
\label{acknowledgments}

C.G. and P.S. have received funding from the European Union FP7/2007-2013 under REA
Grant agreement No. 630925 -- COHEAT and from MIUR-FIRB2013 -- Project Coca (Grant
No.~RBFR1379UX). 
F.G. acknowledges the European Research Council under the European Union's Seventh Framework Program (FP7/2007-2013)/ERC Grant agreement No.~615187-COMANCHE for partial financial support.
\end{acknowledgments}

%\bibliography{biblio}


\end{document}